\documentclass{article}
\usepackage{amsmath}
\usepackage{amssymb}
\usepackage{caption} 
\usepackage{graphicx}
\usepackage{tikz}
\usepackage{multicol}

\title{Probing new physics at the LUXE experiment}
\author{Arka Santra\thanks{for the LUXE Collaboration} \\
	Weizmann Institute of Science\\234, Herzl Street, Rehovot, Israel-7610001  \\
	}

\date{}
\begin{document}

\maketitle

\begin{abstract}

The proposed Laser Und XFEL (LUXE) Experiment at DESY, Hamburg, aims to probe QED in the strong field non-perturbative regime. This regime will be created in collisions between high-intensity laser pulses and high-energy electron or photon beams coming from the Eu.XFEL setup. This experiment comes with a unique opportunity to probe physics beyond the standard model. In this proceeding, it is described that with the help of the large photon flux generated at LUXE,  axion-like-particles can be probed up to a mass of 350 MeV and with a photon coupling of $3\times10^{-6}$ GeV$^{-1}$.  This reach in parameter space is comparable to the projected reach of future experiments like FASER2 in the HL-LHC and NA62 in the dump mode.

\end{abstract}

\section{Introduction}
\label{sec:intro}
The standard model (SM) of particle physics is very successful in describing most of the interactions in nature. Yet it does not describe the complete picture of nature, for example dark matter, hierarchy problem and cosmological baryon asymmetry etc. This motivates the search for beyond the SM (BSM) particles/interactions at different energy scales and different kind of experiments. In spite of all the efforts, we are yet to find any ``smoking gun'' of a direct observation of BSM. This suggests that we need to use new experimental techniques which may open alternative ways to explore the uncharted regions.

The non-perturbative and non-linear quantum electrodynamics (QED), such as Schwinger pair production in the presence of a strong electric field~\cite{Sauter,Euler,Schwinger} is still largely unexplored.  The proposed LUXE (Laser Und XFEL Experiment) experiment~\cite{luxe} wants to inspect this strong field QED (SFQED) with the help of the interaction between the high energetic electron beam from the European XFEL (Eu.XFEL) facility and very intense lasers.  An intense laser can be used as an effective optical dump for electrons where electrons emit a large flux of hard photons. Since the photons negligibly interact with the electromagnetic field of the laser, the hard photons can free stream in the laser after production. With an additional forward detector, the LUXE experiment can probe a well-motivated uninvestigated parameter space of a new spin-0 particle (scalar or pseudo-scalar) which couples to photons. This is indicated as LUXE-NPOD: New Physics at Optical Dump~\cite{npod}. In this proceeding, it will be shown that this setup can probe scalar and pseudo-scalar particle masses up to $\mathcal{O}(350)$ MeV and decay constants of $\mathcal{O}(10^5-10^6)$ GeV$^{-1}$.

\section{Probing the New Physics}
\label{sec:newphysics}

The focus of this proceeding is the production of new spin-0 particles motivated by many extensions of SM. The pseudo-scalar particles,  denoted generically as axion-like-particle (ALP), can naturally address the strong CP problem~\cite{CP1,CP2},  address the Heirarchy problem~\cite{Hier1,Hier2},  can be a valid dark matter candidate~\cite{Dark1,Dark2, Dark3}.  The effective ALP ($a$) interaction with photons and electrons are given by the following Lagrangian:
\begin{equation}
\mathcal{L}_a = \frac{a}{4\Lambda_a}F_{\mu\nu}\tilde{F}^{\mu\nu} + ig_{ae}a\bar{e}\gamma^5e,\label{eq:laga}
\end{equation} 
where $\tilde{F}_{\mu\nu} = \frac{1}{2}\epsilon_{\mu\nu\alpha\beta}F^{\alpha\beta}$,  $g_{ae}$ is the coupling between ALP and electrons and $\Lambda_a$ is the scale of interaction between ALP and the SM particles.  For this discussion, we take $g_{ae}$ to be 0, so ALPs can decay only to photons. The decay rate of the ALP into two photons is shown in the following equation:
\begin{equation}
\Gamma_{a\rightarrow 2\gamma} = \frac{m_a^3}{64\pi\Lambda_a^2}.\label{eq:decay}
\end{equation}
The interaction of the scalar particle ($\phi$) with the photons and electrons can be written in a similar fashion~\cite{npod}.  




\subsection{The Concept of Optical Dump}
\label{sec:opticaldump}


In the LUXE experiment, high flux of free streaming hard photons emitted from the electron-laser interaction region will be used to search for the weakly interacting spin-0 particles; this is the so-called optical dump. This simply refers to the use of the laser as an effective dump for the incoming electrons. In order to successfully meet the criteria of the optical dump where electrons will emit free streaming hard photons, the time scales of the problem should have the following relation:
\begin{equation}
\omega_L^{-1} \ll \tau_\gamma \lesssim t_L \ll \tau_{ee}, \label{eq:timerelation}
\end{equation}
where $\omega_L$ is the laser angular frequency,  $\tau_\gamma$ is the typical timescale of the high intensity Compton scattering,  $t_L$ is the laser pulse length and $\tau_{ee}$ is the typical pair production time scale. These numbers relevant to the LUXE experiment are as follows:
\begin{itemize}
\item the laser wavelength will be 800 nm; that makes $\omega_L^{-1}\sim 0.4$ fs~\cite{luxe},
\item the typical timescale of non-linear Compton scattering is $\mathcal{O}(10)$ fs,
\item the laser pulse length for the LUXE experiment is expected to be $\sim\mathcal{O}(10-200)$ fs,
\item the pair production timescale relevant for the LUXE experiment is $\mathcal{O}(10^4-10^6)$ fs~\cite{luxe}.
\end{itemize}
Hence the relation~(\ref{eq:timerelation}) will hold for the LUXE setup.  

The $\omega_L^{-1}$ is the shortest timescale in the problem - this allows the laser to be treated as a background field to a leading order.  Since $\tau_{ee}$ is the highest timescale here, the emitted photons can be considered as free streaming. Finally as $\tau_\gamma$ is shorter than $t_L$,  the laser behaves as a thick target for the electrons. The combination of the last two points effectively establishes the LUXE experiment as the source of free hard photons from the electrons, just like a dump but made of lasers, hence optical dump.

\subsection{The LUXE-NPOD proposal}
\label{sec:npodproposal}
\begin{figure}
\centering
\includegraphics[width=0.8\textwidth,angle =0 ]{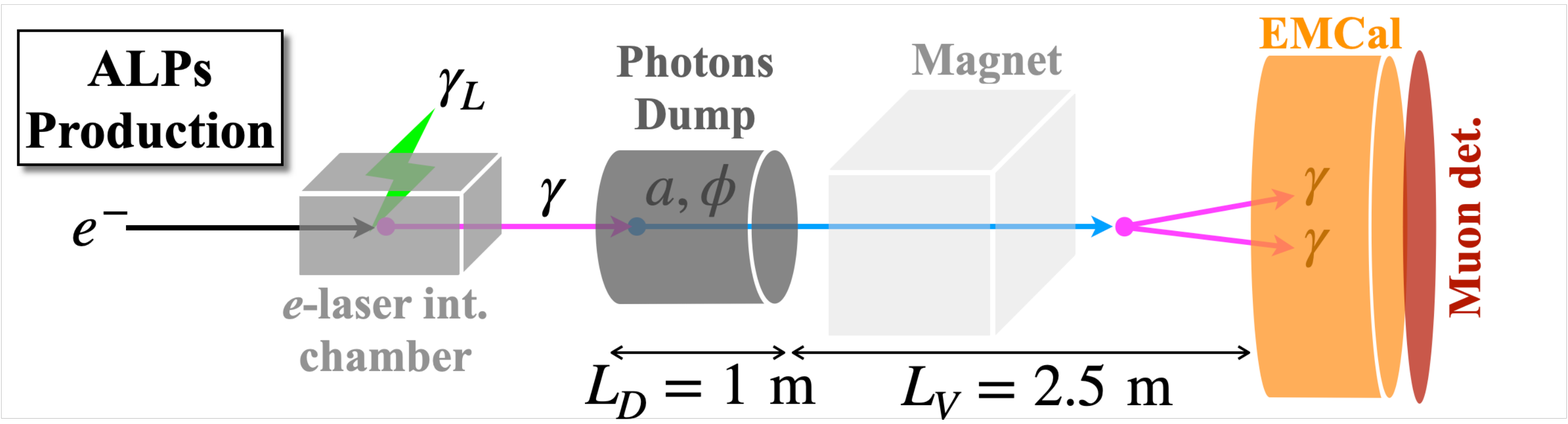}
\caption{A schematic diagram showing the LUXE-NPOD proposal. The free photons streaming from the laser and electron interaction point hit the dump producing the ALPs. The two photons coming from the decay of ALP in the decay volume of $L_V$  are captured by the electromagentic calorimeter kept at the end of the setup.  The muon detector is used to detect the charged muons produced during the SM processes during the shower in the dump.  The magnet in the $L_V$ is kept to bend away the background charged particles and reduce their effect. }
\label{fig:ALP}
\end{figure}

\begin{figure}
\centering
\includegraphics[width=0.42\textwidth, angle =90 ]{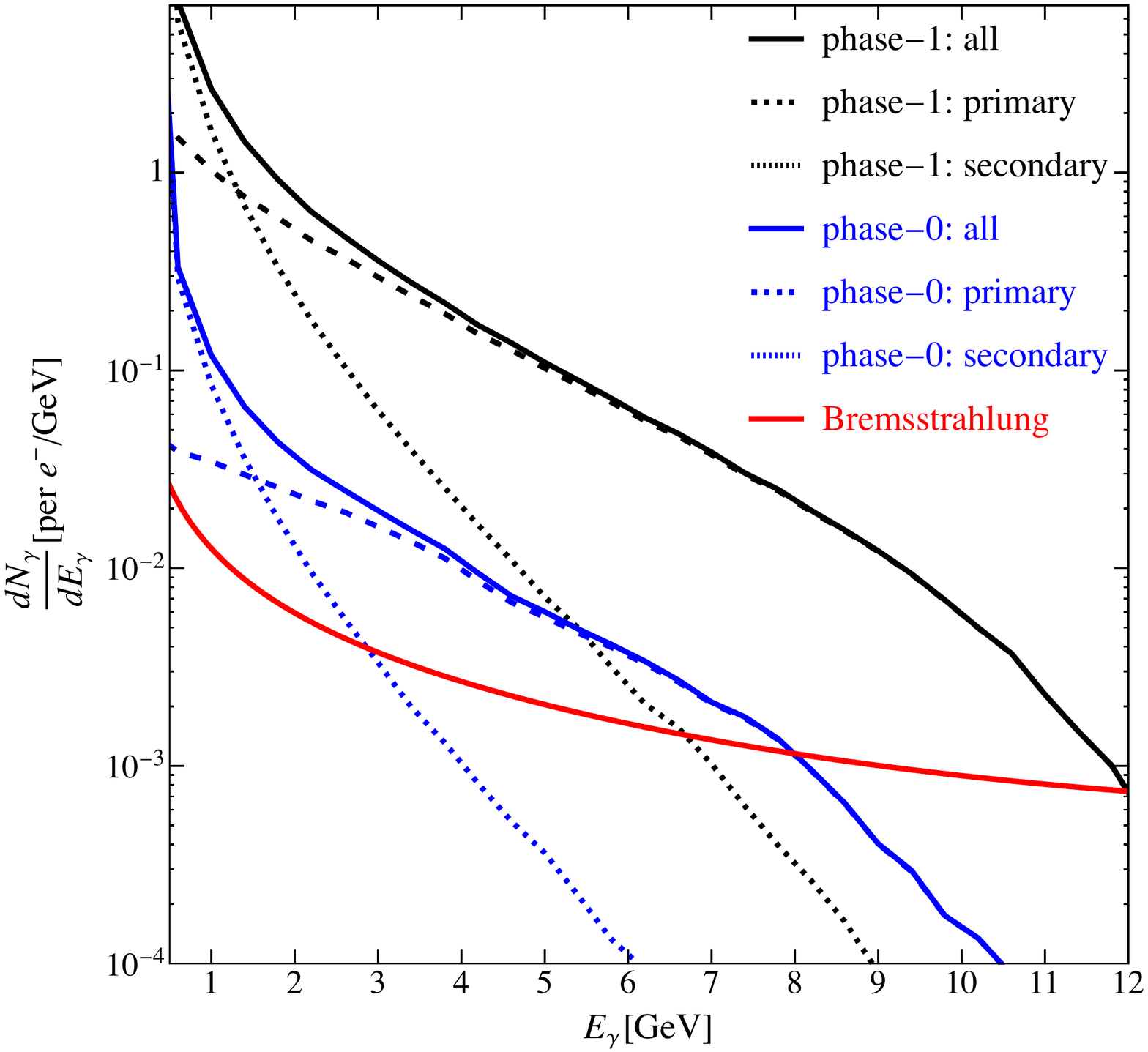}
\caption{The emitted photon spectrum for phase-0 in blue compared with that of the phase-1 in black. The perturbative Bremsstrahlung spectrum with electron energy 16.5 GeV and target length of 0.01 of electron radiation length is shown in red.  In this plot, the primary photons are those which come from the electron-laser interaction point, and the secondary photons come from the shower in the dump.}
\label{fig:inPhoton}
\end{figure}

After the production of free photons in the optical dump, the photons will propagate to a physical dump and interact with the nucleus of the dump to  produce the $X$ particle ($X$ can be a scalar or pseudo-scalar as discussed in Sec.~\ref{sec:newphysics}).  Here we focus on the Primakoff production of X's:
\begin{equation}
\gamma +\ N \rightarrow\ N\ +\ X,\label{eq:primakoff}
\end{equation}
where N is nucleus of the dump and $\gamma$ is the photon coming from the optical dump.  In the proposed setup, the dump length is $L_D$ and the position of it is $\sim$13 m away from the electron-laser interaction point in LUXE.  The long-lived X particles will travel some distance before decaying into $\gamma\gamma$.  Therefore there will be an empty volume of length $L_V$ left after the physical dump where $X$ can decay into $\gamma\gamma$. This two photon signature is the signal for the BSM detector system. The expected number of $X$ particles produced at the proposed setup shown in Fig.~\ref{fig:ALP} is given by~\cite{SeaQuest,MuonBeam,npod}:
\begin{equation}
N_X \thickapprox \mathcal{L}_{eff}\int dE_\gamma\frac{dN_\gamma}{dE_\gamma}\sigma_X(E_\gamma)\left(e^{-\frac{L_D}{L_X}} -e^{-\frac{L_V+L_D}{L_X}}\right)\mathcal{A}.\label{eq:nx}
\end{equation}
In the above equation, $\sigma_X({E_\gamma})$ is the Primakoff cross-section which varies as the square of the nuclear charge $Z^2$ of the dump, $\mathcal{L}_{eff}$ is the effective luminosity, $E_\gamma$ is the incoming photon energy spectrum, $\mathcal{A}$ is the angular acceptance and efficiency of the detector. The differential photon flux per electron, $\frac{dN_\gamma}{dE_\gamma}$ as a function of incoming photon energy spectrum for two different phases of the LUXE experiment is shown in Fig.~\ref{fig:inPhoton}.  The laser intensity will be 40 TW for phase-0 of the run, and this will be increased to 350 TW for the phase-1 of the run. The $L_X\equiv c\tau_X p_X / m_X$ is the propagation length of the X particle with $\tau_X$, $p_X$ and $m_X$ being its proper lifetime,  momentum and mass, respectively.


\section{Results}
\label{sec:results}

\begin{figure}
\centering
\includegraphics[width=0.45\textwidth, angle=-90]{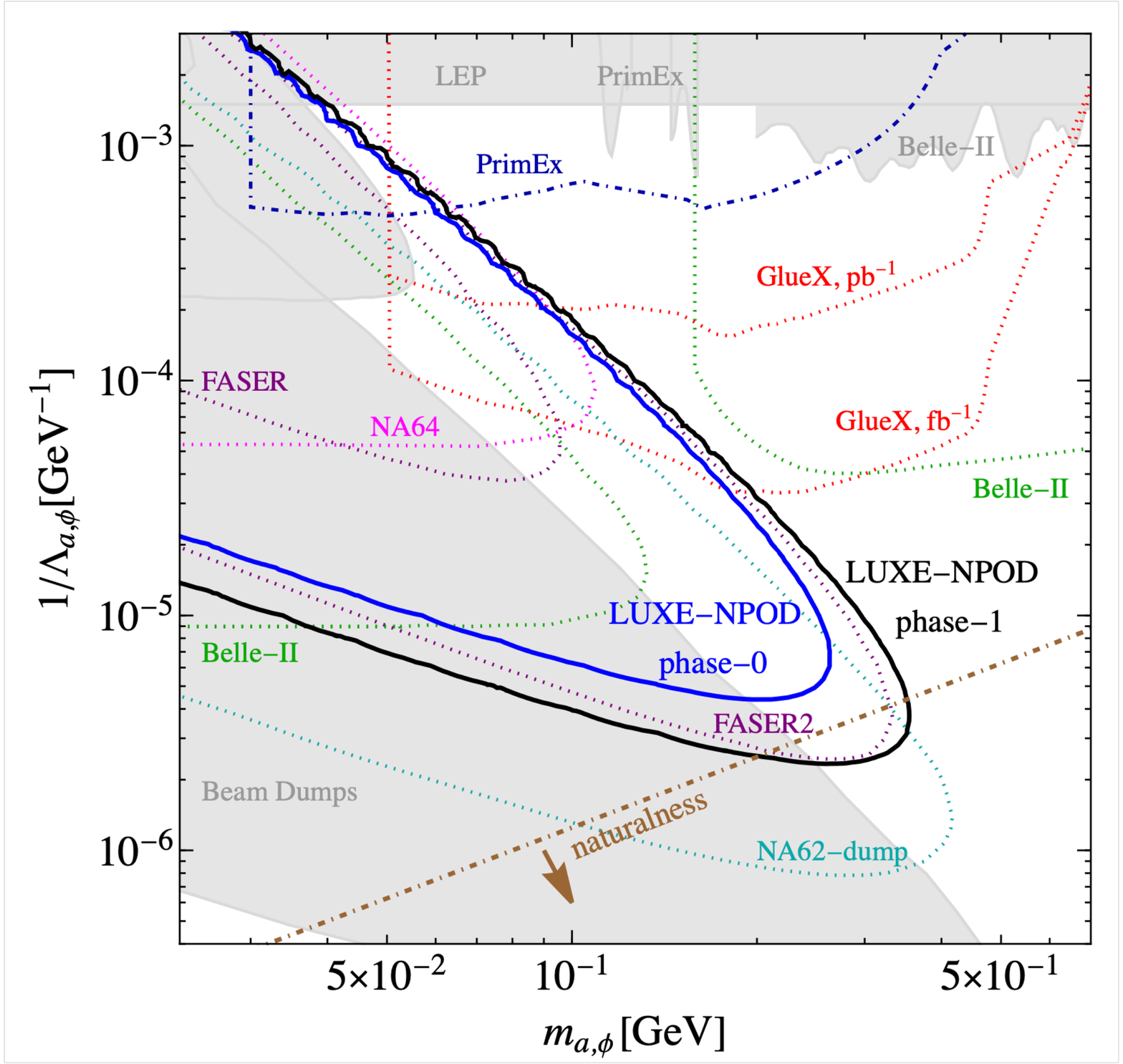}
\caption{The projected reach of the LUXE-NPOD proposal in the effective coupling ($1/\Lambda_X$) vs the mass ($m_X$) parameter space. The phase 0 (1) result is shown in blue (black). The currently existing bounds are shown in grey region. }
\label{fig:exptReach}
\end{figure}

The background for this search consists of (i) charged particles - electrons, muons and hadrons,  (ii) fake photons where neutrons are misidentified as photons and (iii) real photons, from the EM/hadronic interaction close to the end of the dump. The rates of different background particles are estimated using \textsc{Geant4} simulation setup.  The background charged particles can be bent away by a magnetic field of $B\approx1$ T over an active length of $\sim1$ m.  The muon detector will be used to identify muons which may still reach the BSM detector setup. Neutrons can be rejected by looking at their arrival time at the detector because the signal arrives much faster than neutron. The rejection power of the full selection criteria can be kept $\lesssim10^{-3}$.  It is possible to keep neutron to photon fake rate below $10^{-3}$ by utilizing modern detector technology.  For a one year of data taking period ($\sim10^7$ s) and the choice of detector technology, the number of background event is expected to be restricted to less than 1~\cite{npod}. Hence this search is considered effectively background free. The projection of the sensitivity of LUXE-NPOD (effectively the number of expected signal events $N_X=3$, which is the 95\% CL equivalent for background free search) is shown in Fig.~\ref{fig:exptReach} in the effective coupling vs the scalar/pseudoscalar mass plane.  In the phase-0 (phase-1) setup,  LUXE can probe an unexplored region in the mass range from 50 MeV to 250 MeV (40 MeV to 350 MeV) and $1/\Lambda_X > 4\times10^{-6}$ GeV$^{-1}$ ($1/\Lambda_X>2\times10^{-6}$ GeV$^{-1}$). The reach of the LUXE-NPOD is comparable to the projection of other future experiments such as NA62 (in dump mode)~\cite{na62} and FASER2 in the HL-LHC run~\cite{hllhc}.



\section{Conclusion}
\label{sec:conclusion}
In this proceeding, a novel way of searching for feebly interacting spin-0 scalar or pseudoscalar particle has been described.  This search uses free streaming hard photons coming from the collision between high energetic electrons and intense laser pulse - laser pulse is used as an effective dump. This hard photon flux can be directed to a physical dump to produce  weakly interacting spin-0 scalar or pseudoscalar particles. The signature of the BSM particles (two hard photons) can be detected by an EM calorimeter in an effectively background free way.  The BSM detector setup can be used in a parasitic mode in the proposed LUXE experiment at DESY without interfering with the main physics goal of this experiment.

\end{document}